\documentclass[sigconf]{acmart}

\usepackage{algorithmic}
\usepackage{array}
\usepackage{hyperref}
\usepackage{cleveref}
\usepackage[linesnumbered,ruled,vlined]{algorithm2e}
\usepackage{url}
\usepackage{subcaption}

\usepackage{mathtools}

\SetKwInOut{Input}{input}
\SetKwInOut{Output}{output}
\SetNlSty{texttt}{\color{red}}{}
\SetFuncSty{textit}
\SetAlFnt{\footnotesize}
\SetAlgoNlRelativeSize{-2}
\SetKwComment{Comment}{\color{gray}// }{}

\AtBeginDocument{%
  \providecommand\BibTeX{{%
    \normalfont B\kern-0.5em{\scshape i\kern-0.25em b}\kern-0.8em\TeX}}}

\setcopyright{acmcopyright}
\acmDOI{}

\acmConference[SEAMS 2022]{The 17th Symposium on Software Engineering for Adaptive and Self-Managing Systems}{May 21--29, 2022}{Pittsburgh, PA, USA}
\acmISBN{}

\begin{document}

\title{MLProxy: SLA-Aware Reverse Proxy for Machine Learning Inference Serving on Serverless Computing Platforms}


\author{Nima Mahmoudi}
\affiliation{%
  \institution{\textit{Dept. of Electrical and Computer Engineering} \\
  \textit{University of Alberta}}
  \city{Edmonton}
  \state{AB}
  \country{Canada}
}
\email{nmahmoud@ualberta.ca}

\author{Hamzeh Khazaei}
\affiliation{%
  \institution{\textit{Dept. of Electrical Engineering and Computer Science} \\
  \textit{York University}}
  \city{Toronto}
  \state{ON}
  \country{Canada}
}
\email{hkh@yorku.ca}

\renewcommand{\shortauthors}{Mahmoudi and Khazaei, et al.}

\begin{abstract}
  Serving machine learning inference workloads on the cloud is still a challenging task on the production level. Optimal configuration of the inference workload to meet SLA requirements while optimizing the infrastructure costs is highly complicated due to the complex interaction between batch configuration, resource configurations, and variable arrival process. Serverless computing has emerged in recent years to automate most infrastructure management tasks. Workload batching has revealed the potential to improve the response time and cost-effectiveness of machine learning serving workloads. However, it has not yet been supported out of the box by serverless computing platforms. Our experiments have shown that for various machine learning workloads, batching can hugely improve the system's efficiency by reducing the processing overhead per request.

In this work, we present MLProxy, an adaptive reverse proxy to support efficient machine learning serving workloads on serverless computing systems. MLProxy supports adaptive batching to ensure SLA compliance while optimizing serverless costs. We performed rigorous experiments on Knative to demonstrate the effectiveness of MLProxy. We showed that MLProxy could reduce the cost of serverless deployment by up to 92\% while reducing SLA violations by up to 99\% that can be generalized across state-of-the-art model serving frameworks.
\end{abstract}





\maketitle

\section{Introduction}
\label{sec:intro}

Serverless
computing platforms handle almost every aspect of the system administration tasks needed to deploy
a workload on the cloud.
They provide developers with several potential benefits like
handling all of the system administration operations and improving resource utilization, 
leading to potential operational cost savings, improved energy efficiency, and more straightforward application
development~\cite{awsserverless, jonas2019cloud}.

There are typically three phases when using machine learning models in software applications:
1) Model Design, 2) Model Training, and 3) Model Inference (or Model Serving)~\cite{ali2020batch}.
Previous research have found model serving as one of the most challenging
stages in the evolution of the use of machine learning components in commercial
software systems~\cite{lwakatare2019taxonomy}. Some studies have focused on using
different paradigms in cloud computing to improve model serving in terms of latency, throughput,
and cost~\cite{zhang2019mark,crankshaw2020inferline,crankshaw2017clipper,ali2020batch,wu2021serverless}. However, these solutions introduce a higher infrastructure management
overhead compared to managed services or serverless computing platforms.

Serverless computing proves to be a promising option for the deployment of machine learning
serving workloads. Using serverless computing platforms, the developer can only write the code
and leave all infrastructure management tasks to the cloud provider. This paradigm can help the
developer achieve several non-functional goals like low latency and rapid autoscaling while still
being cost-effective. Most serverless providers now can deliver on performance isolation
at any scale, which is very important for applications with unpredictable service request patterns.
Serverless computing can also reduce the cost of deployment because in this paradigm, the developer
is only billed for the actual usage of the resources instead of the provisioned resources. Ease of
management is also another trait of using serverless computing platforms.

For this study, we have leveraged Knative~\cite{knative} on top of Kubernetes~\cite{kubernetes}
as the underlying serverless platform used to perform autoscaling and computations necessary
to serve the model inference workload.
There are several benefits to using Knative and Kubernetes for serving machine
learning inference workloads~\cite{cox2020serverless}.
One of the major benefits of using Kubernetes and Knative for inference
serving is the ability to use autoscaling provided by serverless computing on hardware accelerators
like GPU and TPU.
Using Knative also has the benefit of giving the developers the opportunity to either use
a managed service (e.g., Google Cloud Run) or a self-hosted version
(Knative deployed on the developer's Kubernetes cluster).

Batching has been previously used in other computing paradigms to improve device utilization and cost~\cite{zhang2019mark,crankshaw2020inferline,crankshaw2017clipper,cox2020serverless}.
Batching several requests into a single request can increase the input dimension of the inference,
providing great opportunities for parallelization, especially on accelerated hardware.
However, there are several challenges that need to be overcome
when serving machine learning workloads on serverless computing platforms. Current serverless
computing platforms do not support batching natively. As a result, custom middleware is needed
to allow batching support. Current serverless computing platforms are oblivious to SLA
requirements, while strict SLA requirements are necessary to ensure a good user experience.
To ensure that SLA objectives are met during different arrival rates, adaptive parameter tuning
is necessary~\cite{ali2020batch}.

Many recent studies have focused on optimizing machine learning serving on
AWS Lambda by using profiling and prior access to the deployed models or the
arrival process~\cite{ali2020batch,wu2021serverless}. 
In this work, we strive to build an adaptive system called MLProxy that can
function on both managed and self-hosted serverless platforms without prior
profiling steps using lightweight adaptive batching.
This is necessary for serverless computing platforms
with pay-per-use pricing and allows the developed middleware
to act as a drop-in replacement for API gateways providing instantaneous improvements
over previous methods.

The proposed optimizer has been validated by extensive experimentation on a Knative deployment on an academic
cloud computing infrastructure and works with any workload that can be deployed as Docker containers
and accepts HTTP requests.
Our experiments have shown the effectiveness of MLProxy on several frameworks,
including Tensorflow and Tensorflow Serving, BentoML, PyTorch, and Keras.

Serverless computing is predicted to host most of the future cloud computing workloads~\cite{jonas2019cloud}.
However, the current implementation of serverless computing platforms has not reached its potential
and performs poorly with machine learning inference workloads. In this work, we are trying to introduce
an easy to implement, deploy, and adopt API gateway alternative that improves the performance
of serverless computing platforms in machine learning inference workloads.

The remainder of this paper is organized as follows: 
In \Cref{sec:mlproxy}, we elaborate on the details of MLProxy.
In \Cref{sec:evaluation}, we present the experimental evaluation of the proposed optimizer.
\Cref{sec:results} outlines the experimental results achieved.
In \Cref{sec:related-work}, we survey the latest related work in the optimization of machine learning inference workloads.
\Cref{sec:conc} summarizes our findings and concludes the paper.

\begin{figure*}[htbp]
\centerline{\includegraphics[width=1.3\columnwidth]{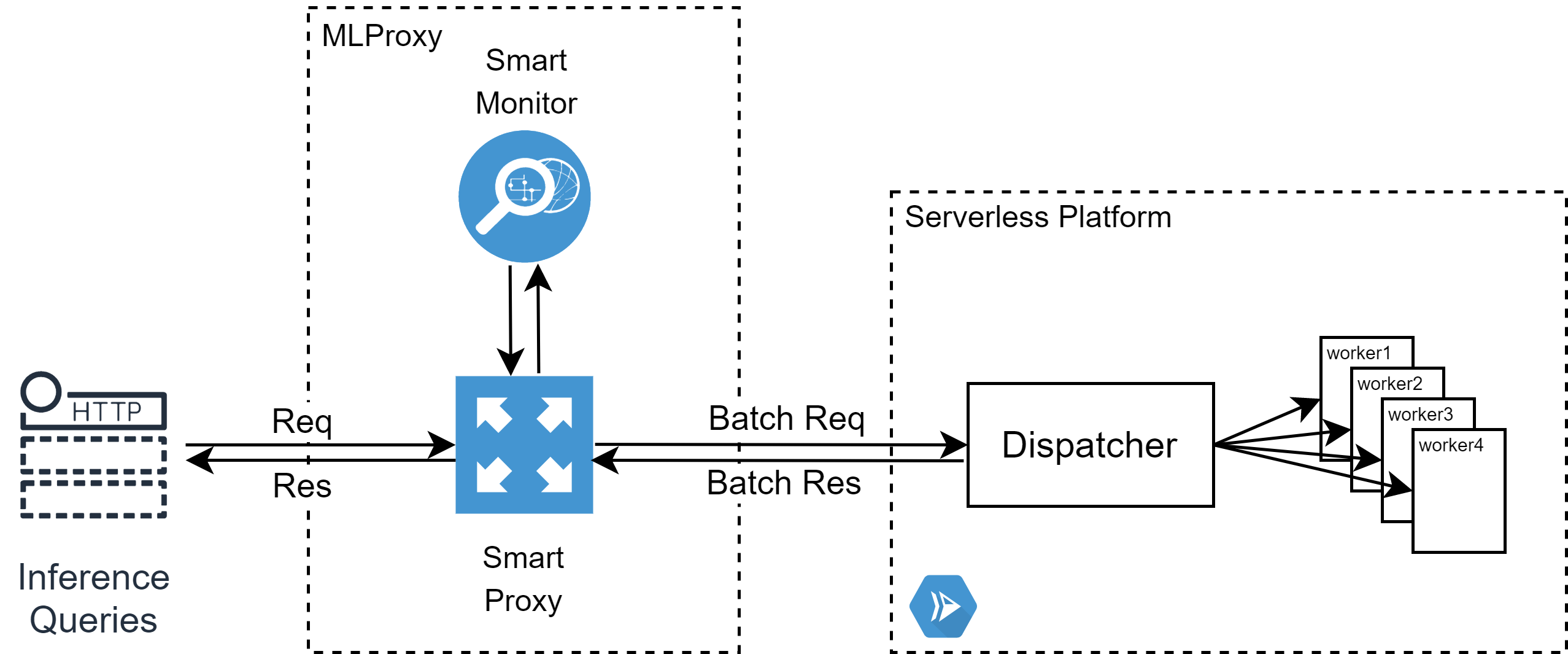}}
\caption{MLProxy Overview.}
\label{fig:mlproxy-overview}
\end{figure*}

\section{ML Proxy} \label{sec:mlproxy}

Given the benefits of batching requests together to improve the utilization and
cost of serverless machine learning deployments, and considering the fact that
the current generation of API gateways do not support batching out of the box, 
we strive to design an SLA-aware
batch optimization platform for these workloads. To facilitate a smooth transition from
the default serverless deployment to our framework, we designed MLProxy as a drop-in
replacement for API gateway systems already existent in every serverless computing
platform, and it will provide instant improvements in cost efficiency while making sure
that the deployment stays SLA-compliant. 
In this section, we will go over our design to address these shortcomings.

\subsection{System Architecture} \label{sec:arch}

\Cref{fig:mlproxy-overview} shows an overview of the MLProxy architecture. As can be seen,
MLProxy acts as an adaptive reverse proxy that can function as a drop-in replacement for the API
Gateway and comprises two modules: 1) Smart Proxy; and 2) Smart Monitor. 

The Smart Proxy module
is designed to accept incoming HTTP requests, group them using our dynamic batching algorithm,
and send them to their respective upstream serverless platform as soon as either the maximum
batch size is reached or the timeout has expired.

In order for the Smart Proxy module to be able to function properly, set accurate timeouts for
batches, and to facilitate accurate dynamic batching, we needed a smart monitoring system
that is tailored to the specific characteristics of machine learning inference workloads
with varying batch sizes. To this end, we designed Smart Monitor as a module that works
alongside the Smart Proxy module and provides several statistical insights regarding
windowed latencies for different batch sizes from the upstream serverless computing platform.

To function properly, MLProxy needs workload
configurations, describing the Service Level Objective (SLO) and endpoints to the upstream serverless platform. Using live
data from the monitoring component, the smart proxy component is able to improve the cost and
reliability of the system by leveraging dynamic batching.

\begin{algorithm}[!ht]
\SetAlgoLined
\KwIn{$Max\_BS$ --- maximum batch size}
\KwIn{$RT\_SLO$ --- response time specified in SLO}
\KwIn{$DTO$ --- Dispath Timeout}
\KwIn{$TO$ --- Timeout}
\KwOut{$metrics$}

 $BS \gets 0$ \Comment*{the size of the current batch}
 $FRT \gets reset$ \Comment*{first request timer}
 \While{True}{
    wait for new arrival or timeout\;
    
    \If{new arrival}{
        cancel previous timeout\;
        \If{BS=0}{
            $FRT \gets reset$
        }
        $BS \gets BS+1$\;
        $P95\_est \gets \text{95th percentile serverless latency for} BS+1$\;
        $DTO \gets RT\_SLO - P95\_est$\;
        $TO \gets DTO - FRT$\;
        \eIf{$BS=Max\_BS$}{
            dispatch current batch to serverless platform\;
            $BS \gets 0$\;
        }{
            set timeout to $TO$\;
        }
    }
    \If{timeout}{
        dispatch current batch to serverless platform\;
        $BS \gets 0$\;
    }
 }
 \caption{An overview of the high-frequency queue scheduler.}
 \label{algorithm-queue-scheduler}
\end{algorithm}

\begin{algorithm}[!ht]
\SetAlgoLined
\KwIn{$RT$ --- current response time}
\KwIn{$TO$ --- current ratio of batches being dispatched due to timeout}
\KwIn{$TO\_thresh$ --- batch timeout threshold set by the user}
\KwIn{$RT\_SLO$ --- response time specified in SLO}

 $inc\_step \gets 1$\;
 $dec\_mult \gets 0.8$\;
 $Max\_BS \gets 1$\;
 \While{True}{
 $RT, TO \gets \text{updated monitoring data}$\;
 $violation \gets False$\;
 \If{$TO > TO\_thresh$ or $RT > RT\_SLO$}{
    $violation \gets True$\;
 }
 \eIf{$violation = True$}{
    $Max\_BS \gets Max\_BS \times dec\_mult$\;
 }{
    $Max\_BS \gets Max\_BS + inc\_step$\;
 }
 wait for 30 seconds\;
 }
 \caption{An overview of the low-frequency dynamic batch optimizer.}
 \label{algorithm-dynamic-batch}
\end{algorithm}

\subsection{Monitoring}

To properly adjust queue size and timeout configurations, MLProxy needs live monitoring data
from the backend serverless computing platform with estimated response times for different
queue sizes. To do so, the proposed monitoring component
organizes observed response time values for batch requests made to the upstream 
serverless computing platform. To ensure we are using the latest latency values, we use
a sliding window to only use the latest response time values in our estimations.

To ensure that the platform is complying with the SLO configuration, the monitoring service
also logs the end-to-end response time observed by the user using a sliding window. Using
this value, the system can make an informed decision about the queuing policy used for the service.
The end-to-end response time includes the time spent processing the request in MLProxy, the queuing
time, and the response time of the upstream serverless computing platform.

\begin{figure}[!ht]
\centerline{\includegraphics[width=1\columnwidth]{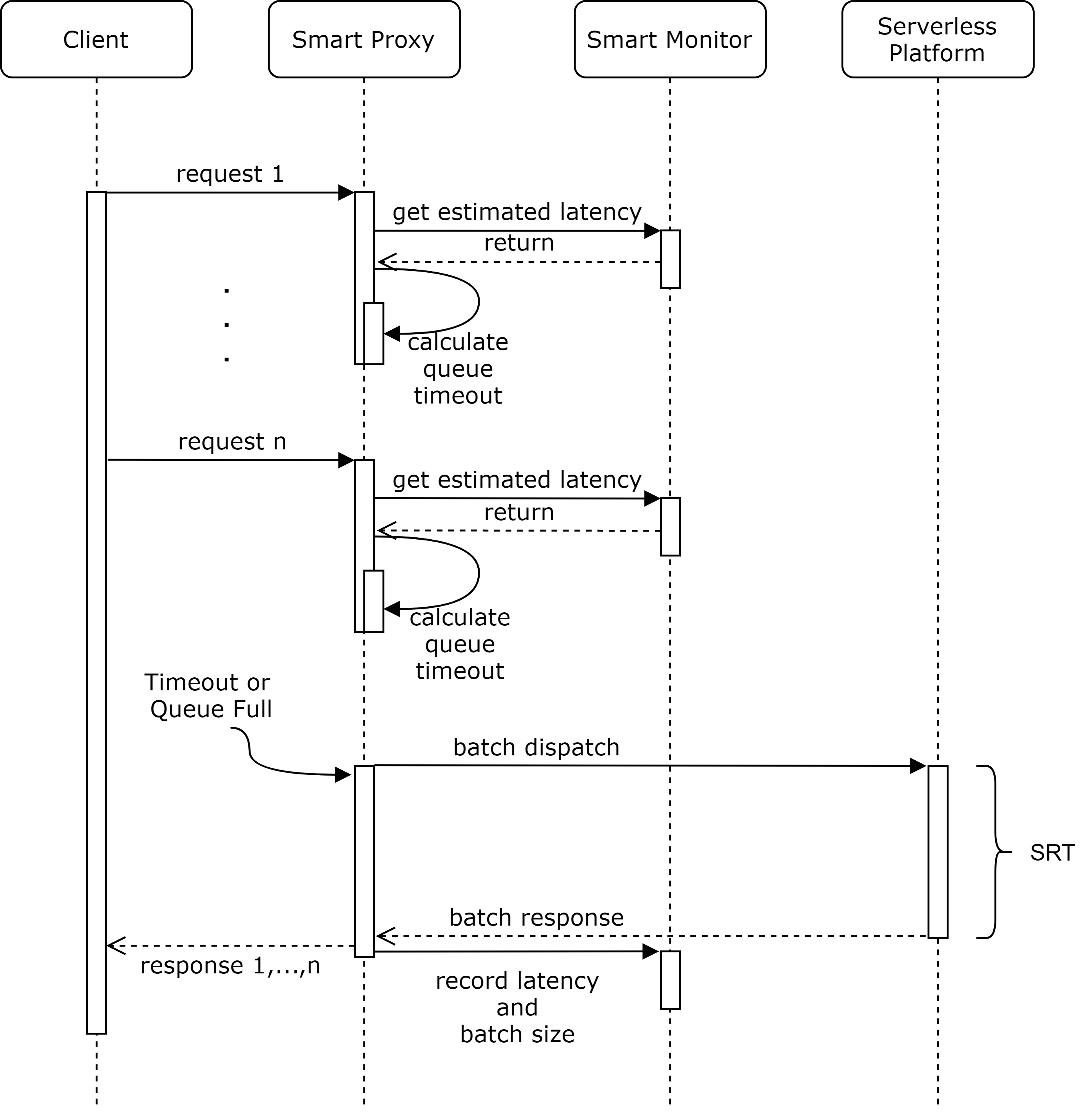}}
\caption{UML Sequence Diagram.}
\label{fig:uml-seq-diagram}
\end{figure}

\subsection{Smart Proxy}

The Smart Proxy component of MLProxy is responsible for creating batches of requests from
individual incoming requests while ensuring end-to-end latency observed by all requests
comply with the set SLO configuration.

\Cref{fig:uml-seq-diagram} shows the UML sequence diagram of a sample batch of requests created
by the Smart Proxy component. Algorithm \ref{algorithm-queue-scheduler} shows the algorithm used
by this component to calculate the timeout and handle the dispatch logic of the batch.
When a request is received by the Smart Proxy, it will be added to the current batch of requests.
There are two triggers that can cause the current batch of requests to be dispatched to the upstream
serverless platform: 1) reaching the maximum batch size; and 2) reaching the queue timeout.
The maximum batch size is set by the dynamic batching subsystem and is responsible for
reducing the cost of deployment while maintaining the quality of service.
The queue timeout is a high-frequency configuration that will be re-evaluated on the arrival
of every request, making sure no request exceeds the latency allowed by the SLO.


\subsection{Dynamic Batch Optimizer}

Batching has been studied by previous work in machine learning inference
as means to improve the resource utilization and throughput of the system~\cite{zhang2019mark,crankshaw2020inferline,crankshaw2017clipper,cox2020serverless}.
Using a larger value for the batch size, we can achieve a lower deployment cost and
higher throughput,
but at the expense of causing a higher latency for the service. The dynamic batching subsystem
is responsible for choosing a batch size that can improve the performance of the system
while ensuring SLA compliance of the system with the current arrival rate observed
by the system.

To achieve the set goals of the system, we adopted an additive increase multiplicative decrease
method for setting the batch size of the system. In this methodology, by default, we will add
a configurable constant amount to the batch size of the system unless we see a violation of
a set of goals for the system. In case of a violation, to ensure the quality of the service
remains within the predefined bounds, we will use a multiplicative decrease on the batch size.
Algorithm \ref{algorithm-dynamic-batch} shows an overview of the logic used by this component.

One of the violations that can cause the scheduler to decrease the batch size is the
latency of the system. As discussed earlier, the SLO requirements of the workload can
be configured by the developer. To ensure the SLO compliance of the workload, we set a lower 
threshold than SLO (80\% of the SLO response time in our experiments) for the system to follow.
Whenever the SLO latency of the system goes beyond this threshold, we trigger a decrease in the
batch size.

Another violation used to decrease the batch size in the proposed scheduler is when we have
too many requests being timed out. This can show that our set batch size is too large for the
current arrival rate, and as a result, the system might introduce very large batch sizes that
can interfere with the functionality of the system.

\subsection{Queuing Scheduler}

\Cref{fig:uml-seq-diagram} shows the UML sequence diagram of a typical
batch processed in the system. 
To ensure the incoming requests are being responded in a
timely fashion that complies with the set SLO, we developed a high-frequency queue timeout
calculator implemented as a part of our smart proxy module. Our timeout calculator calculates
a safe timeout for dispatching requests to the upstream serverless platform in a way that even the
oldest request in the queue is handled before the SLO target latency.

To estimate the overall request latency, we need to estimate the upstream serverless
platform's latency. However, the upstream inference latency depends on the batch size that
is sent to the platform. As a result, the timeout needs to be calculated on each arrival
to adapt to the new queue size ($N_q$). To create such a model, we use the data gathered by 
our monitoring service about the 95th percentile of the latency of the previous 
requests with a batch size of $N_q+1$ to ensure we can fulfill the request before the
deadline, even with the possible arrival of a new request. To calculate the dispatch
timeout, we use the following:

\begin{equation}
    DTO = SLO_T - RT95_{N_q+1}
\end{equation}
where $DTO$ is the dispatch timeout, $SLO_T$ is the Service Level Agreement Target, $RT95_{N_q+1}$
is the 95th percentile of a batch request with $N_q+1$ requests in it,
and $N_q$ is the current batch size.
However, as the timeout will be re-calculated on each arrival, we will set the next timeout
starting from the arrival of the oldest request to make sure all requests will be fulfilled in
time:

\begin{equation}
    TO = DTO - FRT
\end{equation}
where $TO$ is the resulting timeout and $FRT$ (First Request Timer) is the amount of time since the first request in the queue has arrived. Knowing $FRT$ is important to avoid surpassing the SLO threshold. In other words, we need to consider how long the oldest request in the queue has already waited. 

Due to the adaptive nature of the algorithm, $DTO$ might end up as a negative value on each calculation.
In such cases, we dispatch the batch to the upstream serverless platform immediately to avoid SLO violations.

\section{Experimental Evaluation} \label{sec:evaluation}

In this section, we introduce our evaluation of MLProxy using experimentation on our
Knative installation. It is worth noting that the same methodology can be
used on Google Cloud Run, which is a managed Knative offering on the Google Cloud Platform (GCP). The code for performing
and analyzing the experiments used in this section can be found in our public GitHub
repository\footnote{
\url{https://github.com/pacslab/serverless-ml-serving}
}.

\begin{table}[!ht]
\renewcommand{\arraystretch}{1.1}
\begin{center}
    \centering
    \rowcolors{2}{white}{gray!25}
    \caption{Configuration of each VM in the experiments.}
    \label{tab:vm-config}
    \begin{tabular}{l  p{15em}} 
        \hline
        \bf Property & \bf Value\\
        \hline
        vCPU & 8 \\ 
        \hline
        RAM & 30GB \\ 
        \hline
        HDD & 180GB \\ 
        \hline
        Network & 1000Mb/s \\ 
        \hline
        OS & Ubuntu 20.04 \\ 
        \hline
        Latency & \textless 1ms \\ 
        \hline
    \end{tabular}
\end{center}
\end{table}

\subsection{Experimental Setup}

To perform our experiments, we used 4 Virtual Machines (VMs) on the
Compute Canada Arbutus cloud\footnote{\url{https://docs.computecanada.ca/wiki/Cloud_resources}}
with the configuration shown in \Cref{tab:vm-config}.
With this configuration, we were able to utilize 27 vCPU cores for pods running the user's code.
For our cluster, we used Kubernetes version \textit{1.20.5} with 
Kubernetes client (kubectl) version \textit{1.20.0}.
For the client, we used
\textit{Python 3.8.5}. To generate client requests based on a Poisson process,
we used PACSWG workload generation library\footnote{\url{https://github.com/pacslab/pacswg}}
which is publicly available through PyPi\footnote{\url{https://pypi.org/project/pacswg}}.
The result is stored in a CSV file and then processed using
Pandas, Numpy, and Matplotlib.
The dataset, parser, and the code for extraction of system parameters and properties are also
publicly available in the project's GitHub repository.

\begin{table*}[!ht]
\renewcommand{\arraystretch}{1.1}
\begin{center}
    \centering
    \rowcolors{2}{white}{gray!25}
    \caption{List of workloads used in our experiments. The docker container for all workloads along with their code and datasets are publicly available on the project's repository. The baseline response time for each service with 1 vCPU and 1 GB of memory is shown in the complexity column, which shows the complexity of the machine learning model.}
    \label{tab:workloads}
    \begin{tabular}{l l l l l} 
        \hline
        \bf Name & \bf Packaging Lib & \bf ML Lib & \bf Dataset/Model & \bf Complexity \\
        \hline
        SKLearn Iris & BentoML~\cite{bentoml} & Scikit-learn~\cite{scikit-learn} & Scikit-learn~\cite{scikit-learn} & Very Low (8ms) \\
        \hline
        Keras Toxic Comments & BentoML~\cite{bentoml} & Keras/TensorFlow~\cite{keras} & Jigsaw Toxic Comments~\cite{jigsaw-toxic-comments} & Low (40ms) \\
        \hline
        ONNX ResNet50 & BentoML~\cite{bentoml} & ONNX~\cite{onnx} & ONNX Model Zoo~\cite{onnxmodelzoo} & High (201ms) \\
        \hline
        PyTorch Fashion MNIST & BentoML~\cite{bentoml} & PyTorch~\cite{pytorch} & TorchVision Fashion MNIST & Medium (125ms) \\
        \hline
        TFServing MobileNet & TFServing~\cite{tfserving} & TensorFlow~\cite{tensorflow} & MobileNet V1 100x224 & Medium (83ms) \\
        \hline
        TFServing ResNet & TFServing~\cite{tfserving} & TensorFlow~\cite{tensorflow} & ResNet V2 fp32 & High (204ms) \\
        \hline
    \end{tabular}
\end{center}
\end{table*}

\subsection{Workloads}

\Cref{tab:workloads} shows an overview of the workloads used in our experimental studies.
As can be seen, we have experimented on a variety of different machine learning tasks with
a wide range of complexity levels to see how batching would affect these scenarios.

\subsection{Workload Characterization}
In this section, we will go through some workload characterizations that help us understand
our workloads and how we can find workloads that benefit most from our novel dynamic batching.
As the proposed MLProxy uses batching to improve throughput and cost~\cite{crankshaw2017clipper}, there needs to be \textit{some}
benefit in batching requests together for a given workload.
Due to fine-grained billing in serverless computing, if the service time scales linearly with
the batch size, there is no benefit in batching requests together. However,
for workloads with low to medium complexity on CPU and workloads running on accelerated hardware
(e.g., GPU or TPU), service time scales sub-linearly with the batch size,
and thus we can expect improved throughput and deployment cost for these workloads.

\begin{figure}[!ht]
\centerline{\includegraphics[width=1\linewidth]{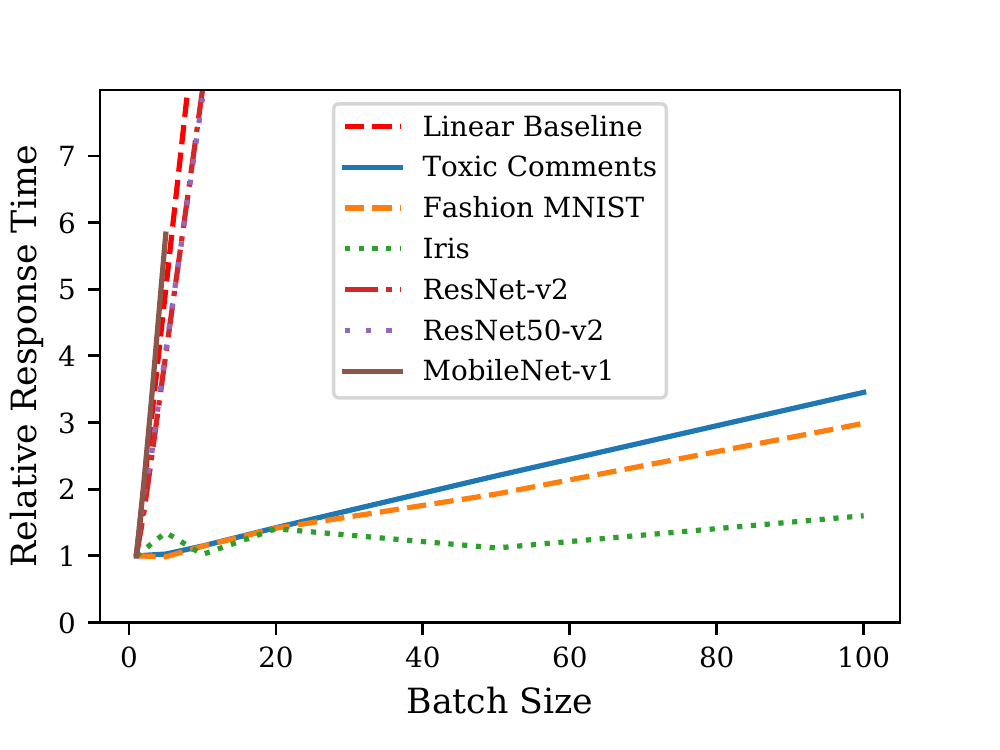}}
\caption{The relative response time against the batch size. Relative response time shows how the average response time grows when increasing the batch size. The linear baseline shows the relative response time that grows perfectly linear with the batch size.}
\label{fig:exp-all-rel-resp-times}
\end{figure}

\begin{figure}[!ht]
\centerline{\includegraphics[width=1\linewidth]{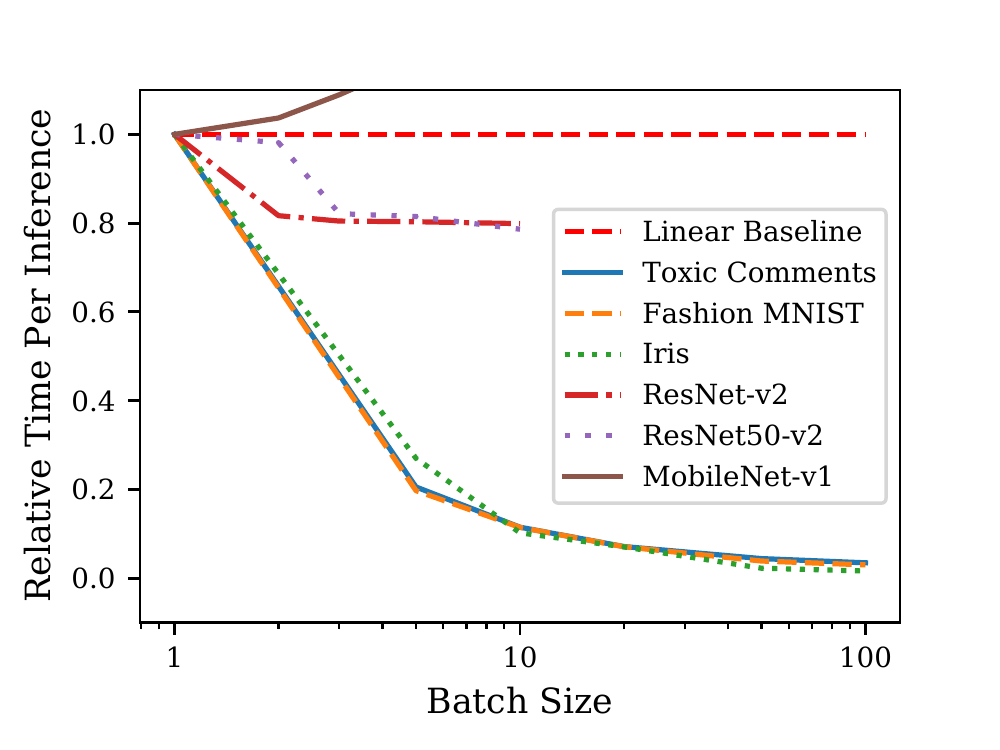}}
\caption{The relative average time per inference against the batch size. For many workloads, increasing the batch size results in a reduction in time spent for each inference by reducing the overhead for each query. For a workload with a response time that grows linearly with the batch size, the time per inference remains the same with any batch size.}
\label{fig:exp-all-inference-time}
\end{figure}

\Cref{fig:exp-all-rel-resp-times,fig:exp-all-inference-time} show how the response time
and time per inference (response time divided by batch size) evolve as we increase the batch size, compared with a batch size of one.
The linear baseline shows a hypothetical workload where the average response time grows linearly
with increasing the batch size. The reason behind this baseline is that in a workload where the average response time
grows linearly with the baseline, the time per inference remains constant for different batch sizes.
Thus, these workloads do not benefit from batching in serverless computing platforms.
As can be seen, in many workloads, time per inference is shorter for larger batch sizes due to lower
computational overhead per inference. Thus, batching queries together can increase the efficiency
and resource utilization of the workload, improving the cost and performance of the deployment.


\subsection{Service-Level Objectives (SLOs)}

Service-Level Agreements (SLAs) are defined around a specific service and serve the purpose of forming an agreement
between the client and the provider of the service, laying out the metrics by which the
service is measured and the penalties if the service level agreed upon is not reached.
Service-Level Objectives (SLOs) serve as the target for a given service metric, e.g., average and 95th percentile of the response time.
SLAs are a more preferred and far more accurate way of representing the needs of the client
from the service as they list out the most important criteria for the client.

In this work, our goal is to improve the cost and throughput of serving machine learning
workloads on serverless computing platforms while ensuring the agreed-upon level of service
is satisfied. For this purpose, we used one of the most common metrics in SLOs, which is the 95th
percentile of the response time. However, both the percentile and the threshold used are configurable
in our proposed MLProxy component.

\subsection{Real-World Traces}
\begin{figure*}[!ht]
\centering
\begin{subfigure}{.33\textwidth}
  \centering
  \includegraphics[width=1\linewidth]{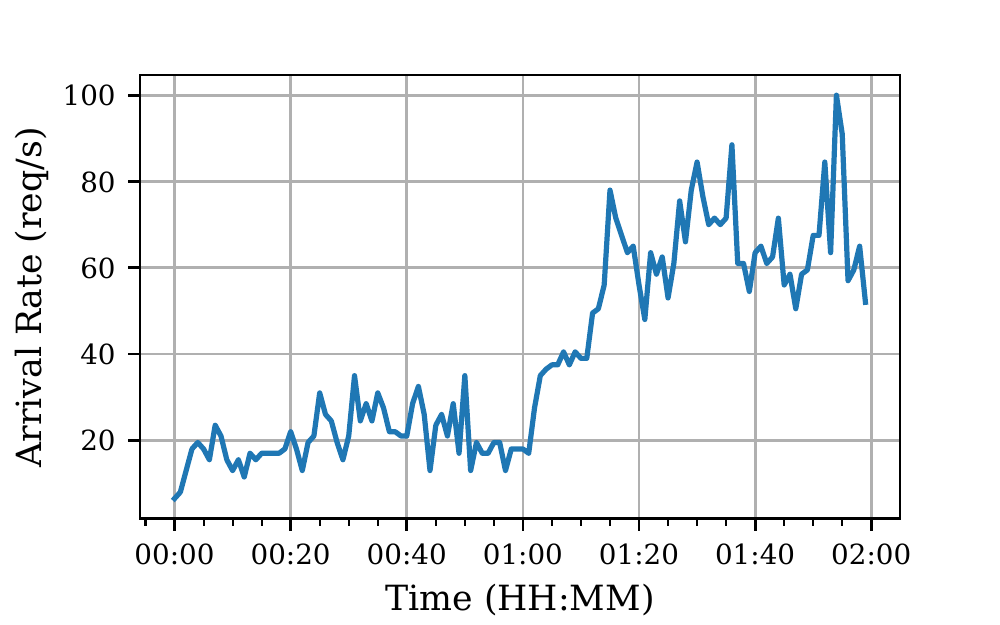}
  \caption{AutoScale NLAR T4 Trace.}
  \label{fig:all-traces-sub1}
\end{subfigure} \hfill
\begin{subfigure}{.33\textwidth}
  \centering
  \includegraphics[width=1\linewidth]{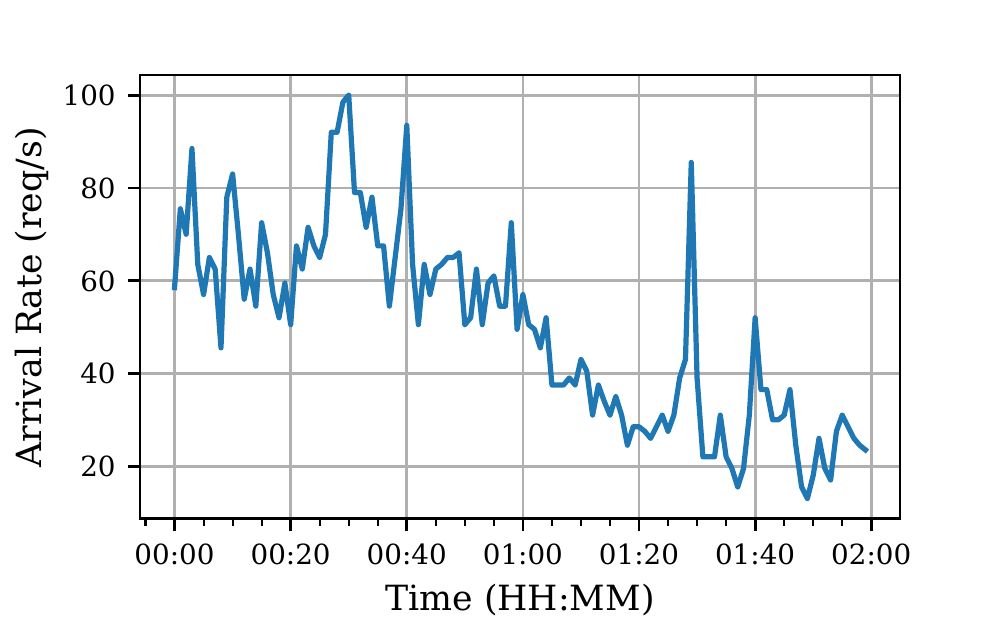}
  \caption{AutoScale NLAR T5 Trace.}
  \label{fig:all-traces-sub2}
\end{subfigure} \hfill
\begin{subfigure}{.33\textwidth}
  \centering
  \includegraphics[width=1\linewidth]{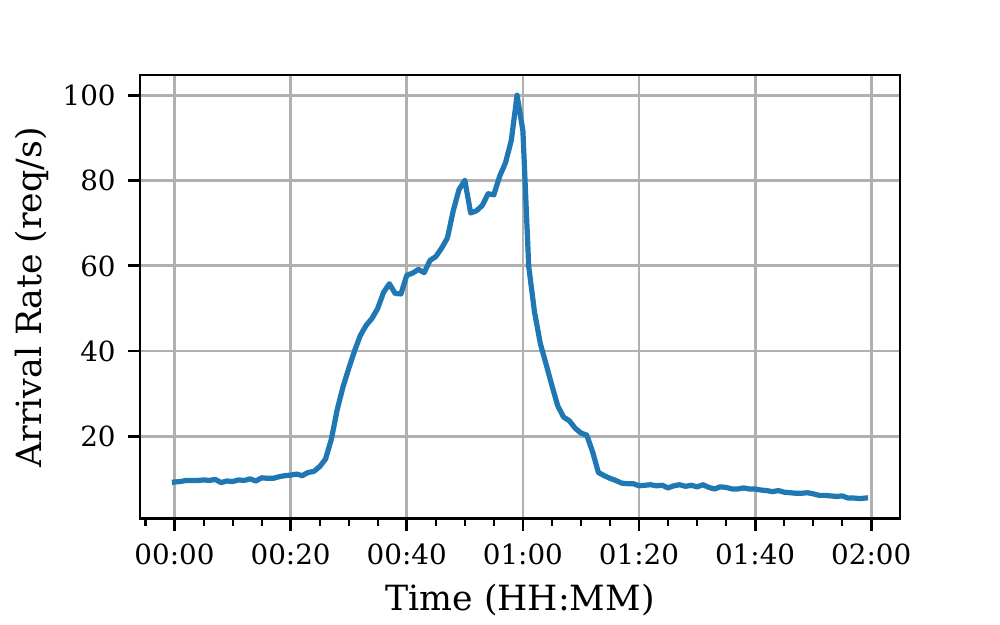}
  \caption{AutoScale FIFA World Cup Trace.}
  \label{fig:all-traces-sub3}
\end{subfigure}
\caption{The trace patterns used in the experiments~\cite{gandhi2012autoscale}.}
\label{fig:all-traces}
\end{figure*}

To emulate real-world scenarios, we have used the AutoScale real-world
traces NLAR T4 and T5 and FIFA World Cup~\cite{gandhi2012autoscale}.
However, to match the capacity of our cluster and to imitate different load
intensities, we scaled the maximum arrival rate for our experiments.
\Cref{fig:all-traces} shows these traces scaled to have a maximum arrival rate of
100 requests per second.

\begin{table*}[!ht]
\renewcommand{\arraystretch}{1.1}
\begin{center}
    \centering
    \rowcolors{2}{white}{gray!25}
    \caption{Experimental results: BRT shows the baseline average response time with a concurrency of one and batch size of one and SLO P95 is the 95th percentile response time specified in SLO. Number of containers is shown as the most important deployment cost indicator. Note that the columns specified with an asterisk (*) are the results with MLProxy turned on.}
    \label{tab:experiments}
    \begin{tabular}{l l l p{2em} p{2em} p{2em} p{2em} p{6em} p{2.5em} p{7em} l} 
        \hline
        \bf \# & \bf Workload & \bf Trace & \bf Max RPS & \bf BRT (ms) & \bf SLO P95 (ms) & \bf \# of Cont. & \bf \# of Cont.* & \bf \% of SLO Viol. & \bf \% of SLO Viol.* & \bf Avg. BS \\
        \hline
        1 & Fashion MNIST & WC & 30 & 125 & 500 & 2.73 & 1.00 ($\downarrow$63.4\%) & 1.2799 & 0.1861 ($\downarrow$85.5\%) & 4.93 \\
        \hline
        2 & Fashion MNIST & WC & 100 & 125 & 1000 & 8.75 & 1.01 ($\downarrow$88.5\%) & 26.0048 & 0.0767 ($\downarrow$99.7\%) & 10.93 \\
        \hline
        3 & Iris & WC & 50 & 8 & 500 & 1.61 & 1.00 ($\downarrow$38.1\%) & 0.8892 & 0.0033 ($\downarrow$99.6\%) & 5.01 \\
        \hline
        4 & Iris & WC & 185 & 8 & 200 & 1.50 & 1.01 ($\downarrow$32.8\%) & 0.2862 & 0.0395 ($\downarrow$86.2\%) & 6.57 \\
        \hline
        5 & Toxic Comments & WC & 30 & 40 & 500 & 1.90 & 1.00 ($\downarrow$47.2\%) & 0.4181 & 0.0811 ($\downarrow$80.6\%) & 3.09 \\
        \hline
        6 & Fashion MNIST & T5 & 30 & 125 & 500 & 4.28 & 1.00 ($\downarrow$76.6\%) & 1.9688 & 0.1002 ($\downarrow$94.9\%) & 9.81 \\
        \hline
        7 & Iris & T5 & 185 & 8 & 500 & 3.01 & 1.00 ($\downarrow$66.7\%) & 0.6675 & 0.0059 ($\downarrow$99.1\%) & 18.95 \\
        \hline
        8 & Iris & T5 & 185 & 8 & 200 & 3.01 & 1.00 ($\downarrow$66.7\%) & 0.7064 & 0.0019 ($\downarrow$99.7\%) & 11.00 \\
        \hline
        9 & Toxic Comments & T5 & 50 & 40 & 500 & 3.87 & 1.00 ($\downarrow$74.2\%) & 0.4771 & 0.0553 ($\downarrow$74.2\%) & 7.71 \\
        \hline
        10 & Fashion MNIST & T4 & 100 & 125 & 1000 & 13.34 & 1.07 ($\downarrow$92.0\%) & 39.9915 & 0.0038 ($\downarrow$99.9\%) & 13.34 \\
        \hline
        11 & Iris & T4 & 185 & 8 & 200 & 1.93 & 1.00 ($\downarrow$48.3\%) & 0.5361 & 0.0295 ($\downarrow$94.5\%) & 13.06 \\
        \hline
        12 & Toxic Comments & T4 & 50 & 40 & 500 & 3.12 & 1.00 ($\downarrow$67.9\%) & 0.4737 & 0.0405 ($\downarrow$91.4\%) & 6.12 \\
        \hline
    \end{tabular}
\end{center}
\end{table*}

\begin{figure*}[!ht]
\centering
\begin{subfigure}{.33\textwidth}
  \centering
  \includegraphics[width=1\linewidth]{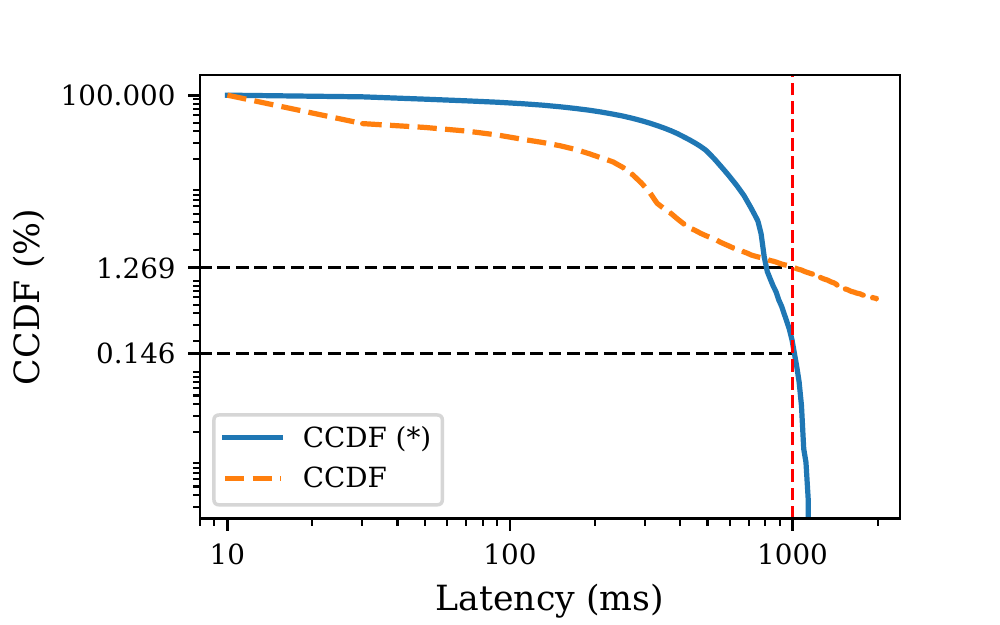}
  \caption{Experiment \#1.}
  \label{fig:all-res-ccdf-sub1}
\end{subfigure} \hfill
\begin{subfigure}{.33\textwidth}
  \centering
  \includegraphics[width=1\linewidth]{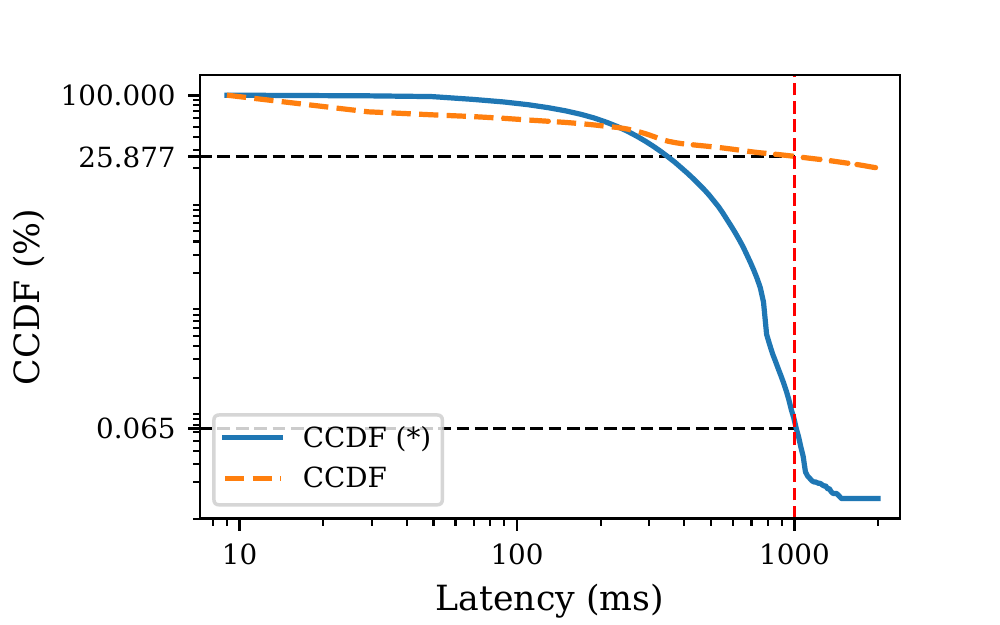}
  \caption{Experiment \#2.}
  \label{fig:all-res-ccdf-sub2}
\end{subfigure} \hfill
\begin{subfigure}{.33\textwidth}
  \centering
  \includegraphics[width=1\linewidth]{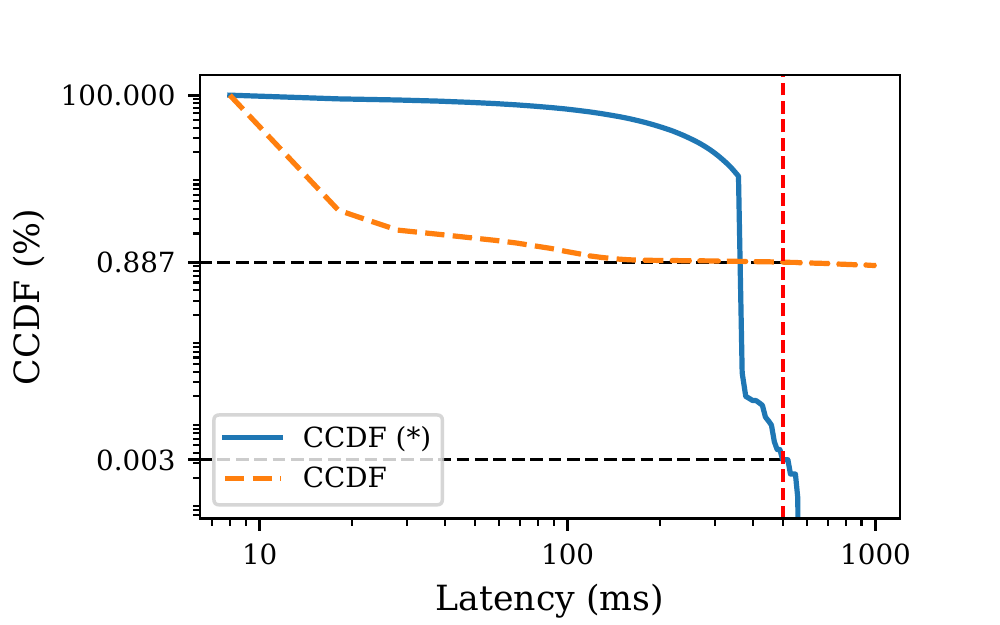}
  \caption{Experiment \#3.}
  \label{fig:all-res-ccdf-sub3}
\end{subfigure}
\begin{subfigure}{.33\textwidth}
  \centering
  \includegraphics[width=1\linewidth]{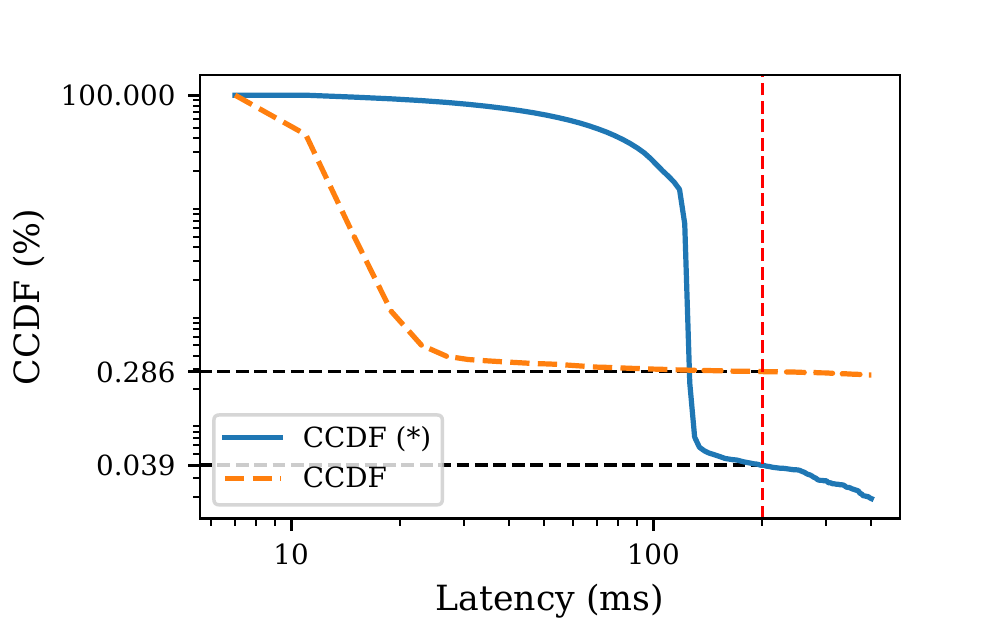}
  \caption{Experiment \#4.}
  \label{fig:all-res-ccdf-sub4}
\end{subfigure} \hfill
\begin{subfigure}{.33\textwidth}
  \centering
  \includegraphics[width=1\linewidth]{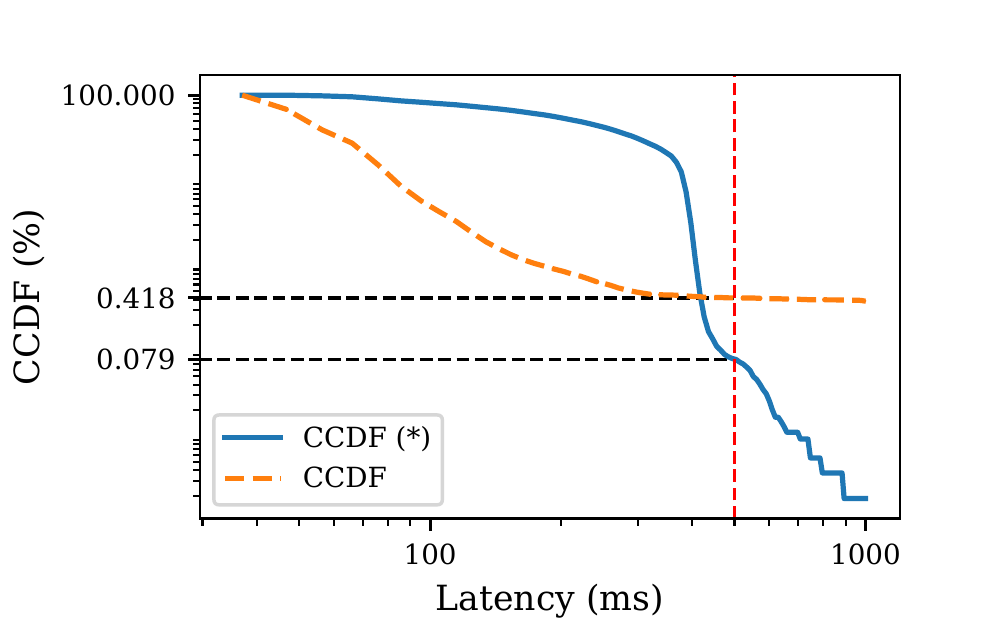}
  \caption{Experiment \#5.}
  \label{fig:all-res-ccdf-sub5}
\end{subfigure} \hfill
\begin{subfigure}{.33\textwidth}
  \centering
  \includegraphics[width=1\linewidth]{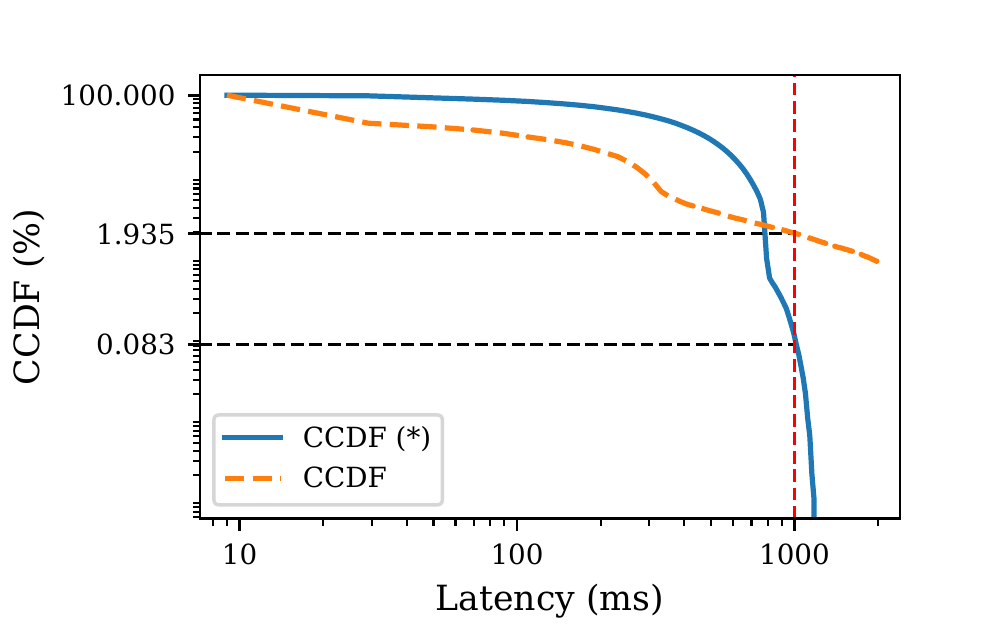}
  \caption{Experiment \#6.}
  \label{fig:all-res-ccdf-sub6}
\end{subfigure}
\begin{subfigure}{.33\textwidth}
  \centering
  \includegraphics[width=1\linewidth]{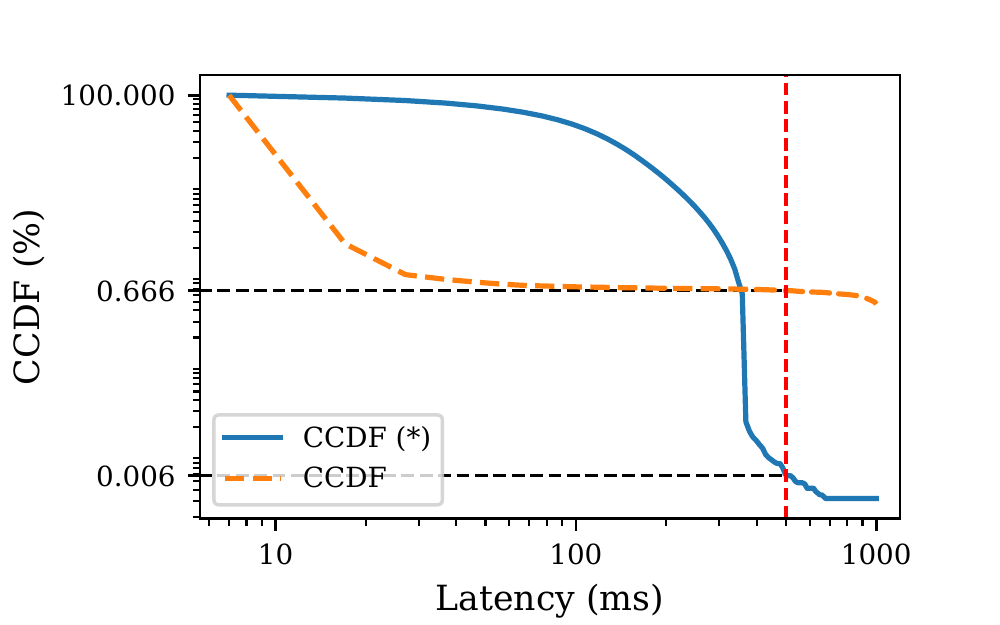}
  \caption{Experiment \#7.}
  \label{fig:all-res-ccdf-sub7}
\end{subfigure} \hfill
\begin{subfigure}{.33\textwidth}
  \centering
  \includegraphics[width=1\linewidth]{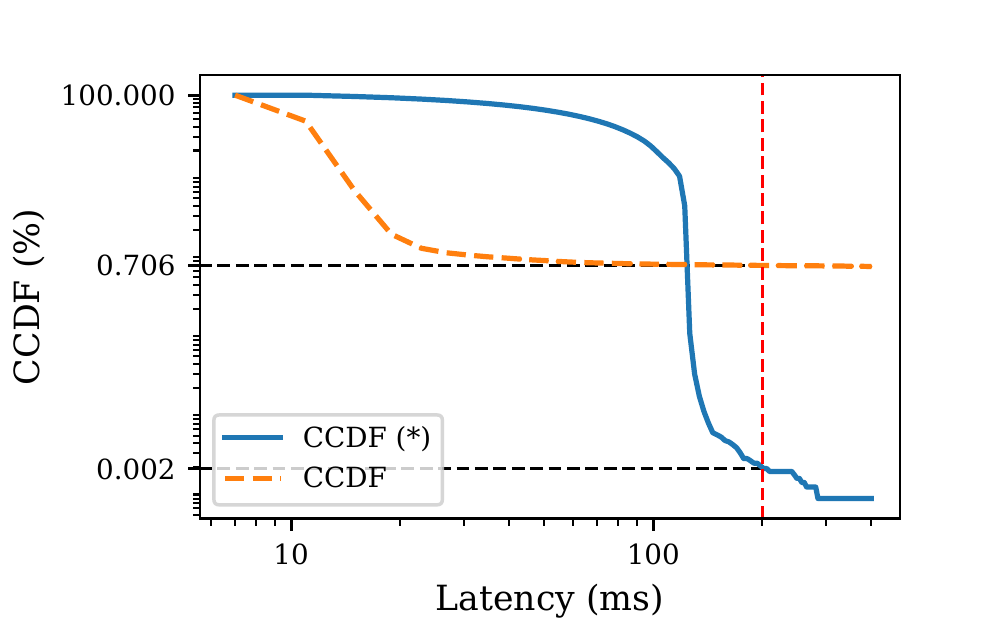}
  \caption{Experiment \#8.}
  \label{fig:all-res-ccdf-sub8}
\end{subfigure} \hfill
\begin{subfigure}{.33\textwidth}
  \centering
  \includegraphics[width=1\linewidth]{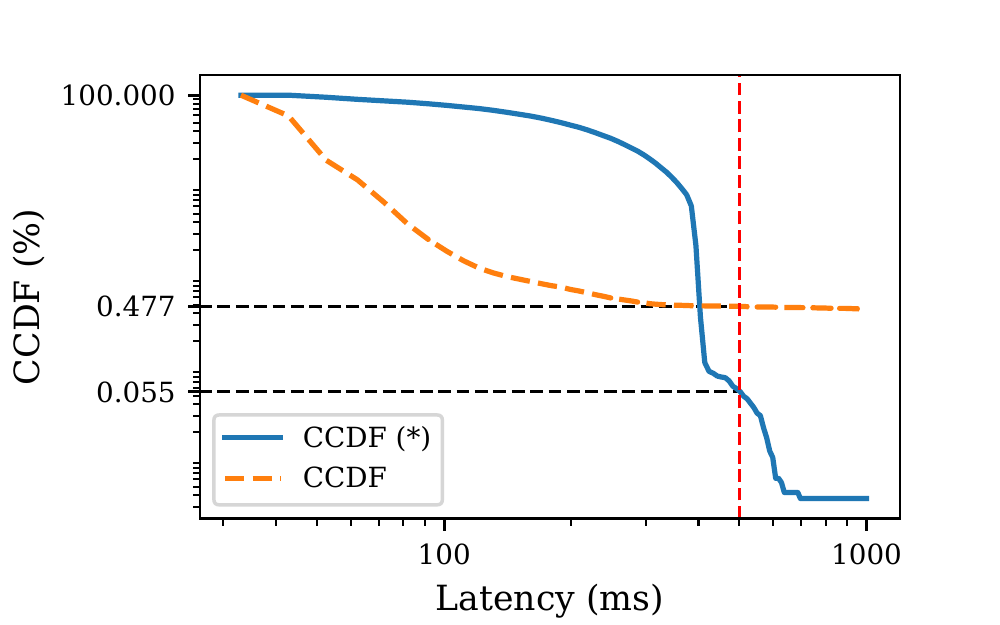}
  \caption{Experiment \#9.}
  \label{fig:all-res-ccdf-sub9}
\end{subfigure}
\begin{subfigure}{.33\textwidth}
  \centering
  \includegraphics[width=1\linewidth]{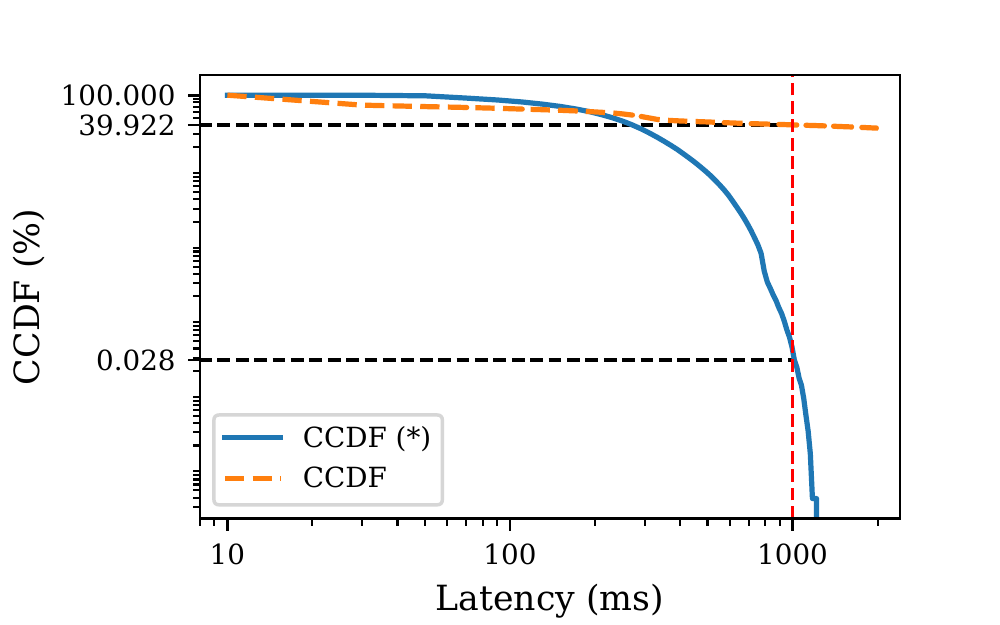}
  \caption{Experiment \#10.}
  \label{fig:all-res-ccdf-sub10}
\end{subfigure} \hfill
\begin{subfigure}{.33\textwidth}
  \centering
  \includegraphics[width=1\linewidth]{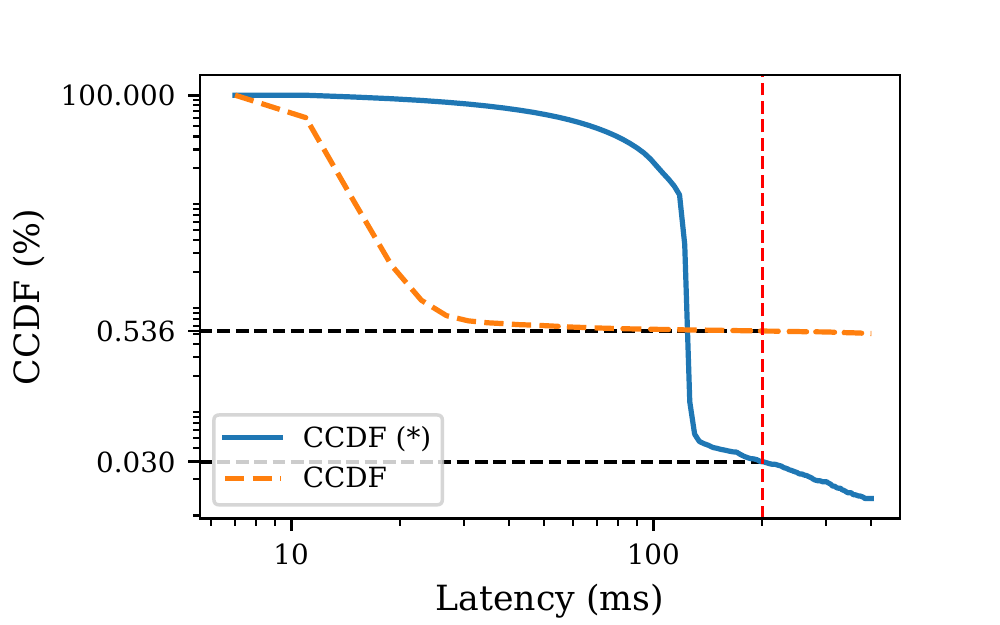}
  \caption{Experiment \#11.}
  \label{fig:all-res-ccdf-sub11}
\end{subfigure} \hfill
\begin{subfigure}{.33\textwidth}
  \centering
  \includegraphics[width=1\linewidth]{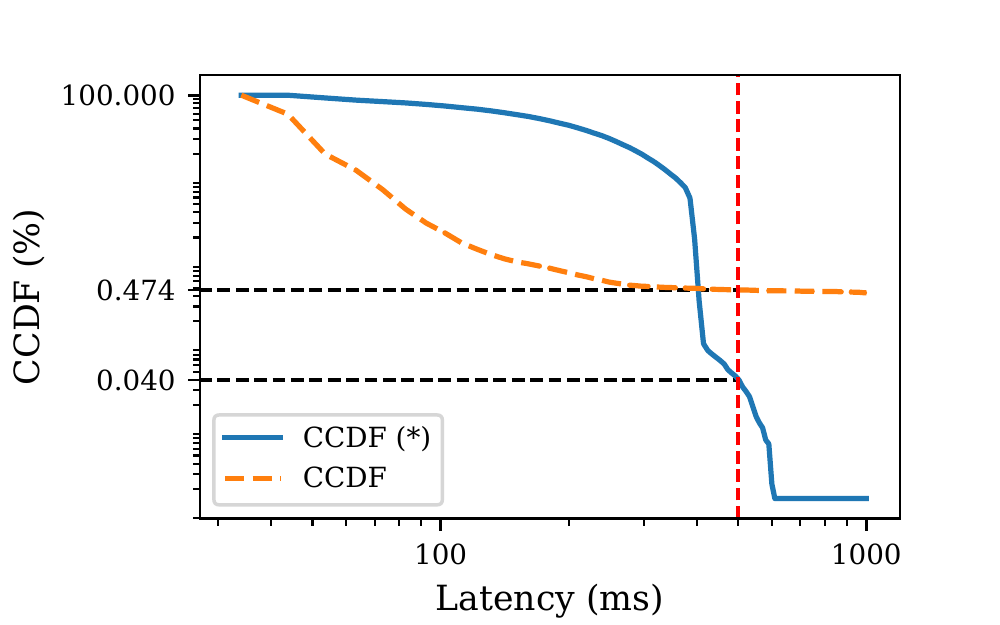}
  \caption{Experiment \#12.}
  \label{fig:all-res-ccdf-sub12}
\end{subfigure}
\caption{The Complementary CDF (CCDF) results of experiments listed in \Cref{tab:experiments}. The red dashed horizontal line shows the SLO set for the experiment and the vertical bars signify the total SLO miss rate of experiments with and without MLProxy optimizer. Plots marked with an asterisk (*) are the results with MLProxy turned on.}
\label{fig:all-res-ccdf}
\end{figure*}

\begin{figure*}[!ht]
\centering
\begin{subfigure}{.45\textwidth}
  \centering
  \includegraphics[width=1\linewidth]{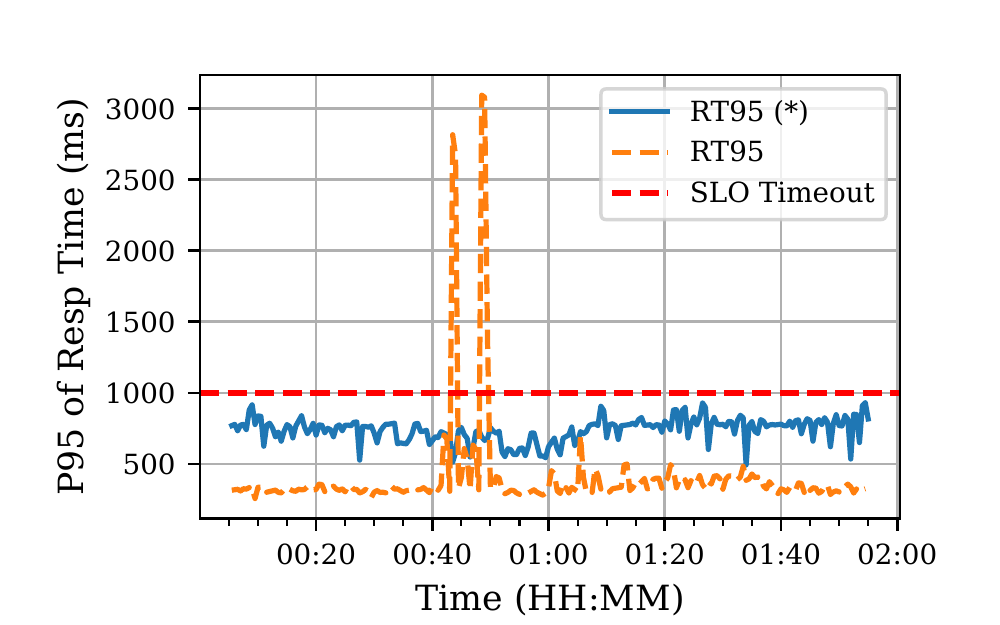}
  \caption{The 95th percentile of the response time over time throughout the experiment.}
  \label{fig:all-res-sub1}
\end{subfigure}\hfill%
\begin{subfigure}{.45\textwidth}
  \centering
  \includegraphics[width=1\linewidth]{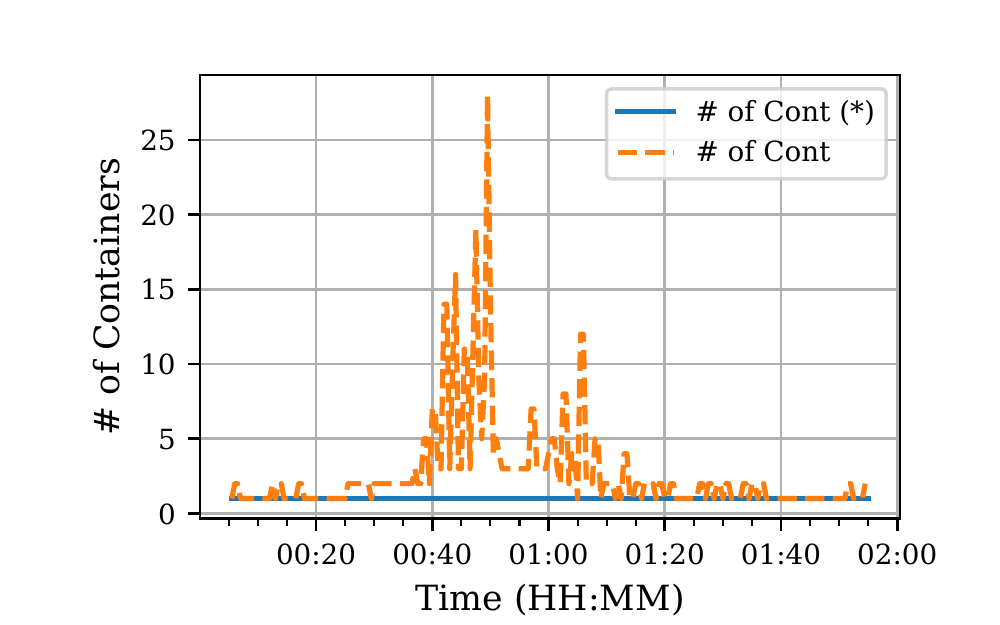}
  \caption{Number of containers over time over time throughout the experiment.}
  \label{fig:all-res-sub2}
\end{subfigure}
\begin{subfigure}{.45\textwidth}
  \centering
  \includegraphics[width=1\linewidth]{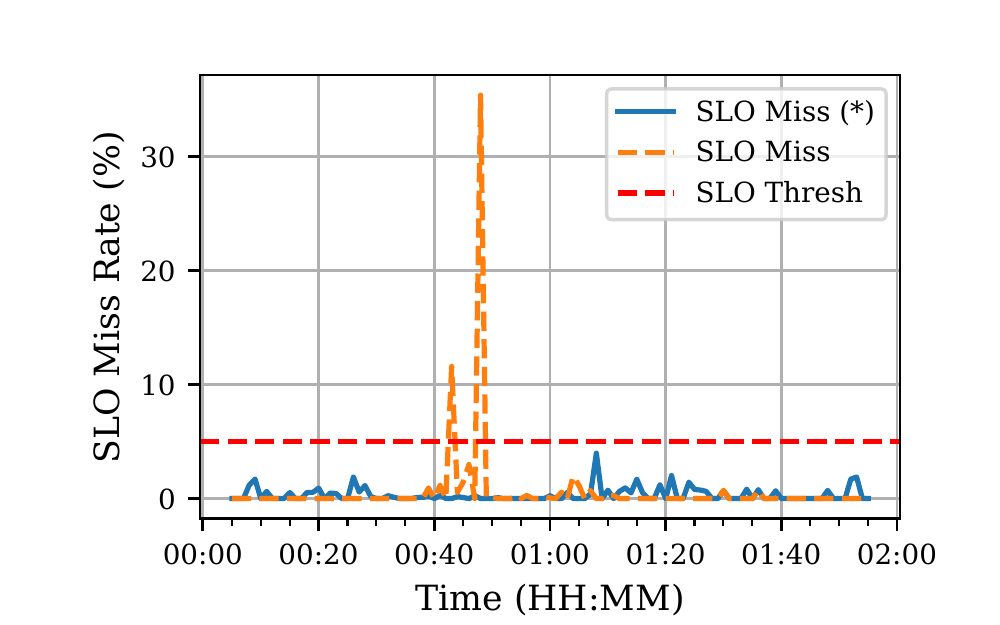}
  \caption{The SLO miss rate over time throughout the experiment.}
  \label{fig:all-res-sub3}
\end{subfigure}\hfill%
\begin{subfigure}{.45\textwidth}
  \centering
  \includegraphics[width=1\linewidth]{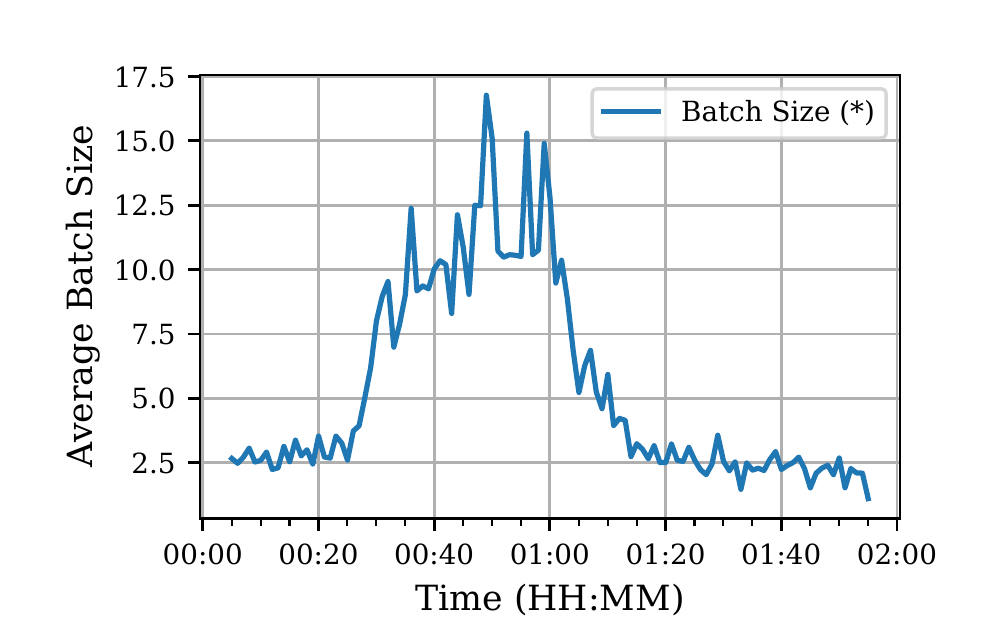}
  \caption{Average batch size over time throughout the experiment.}
  \label{fig:all-res-sub4}
\end{subfigure}
\caption{The experimental results when applying the world cup trace to the Fashion MNIST workload with a scaled maximum arrival rate of 30 requests per second and 95th percentile of response time SLO set to $500ms$ on the optimizer engine. Plots marked with an asterisk (*) are the results with MLProxy turned on.}
\label{fig:all-res}
\end{figure*}

\section{Experimental Results and Discussion} \label{sec:results}

In this section, we will go through our experimental results and discuss them in detail. Please
note that the codes and scripts used to deploy and experiment with workloads and analyze the results
is publicly accessible in the project's GitHub repository\footnote{
\url{https://github.com/pacslab/serverless-ml-serving}
}.


\Cref{tab:experiments} lists a large number of experiments we have performed on our cluster
to determine the efficiency of our method.
Out of the workloads listed in \Cref{tab:workloads}, we have only done further experimentation on the
ones that would benefit from batching.
To further show details of the response times achieved in our experiments with and without 
using the proposed method, the Complementary Cumulative Density Function (CCDF) plot of the
experiments listed in \Cref{tab:experiments} have been shown in \Cref{fig:all-res-ccdf}.
We have also included other plots about the details of experiment \#1 in \Cref{fig:all-res}.

\subsection{SLO Compliance}

As shown in \Cref{tab:experiments}, MLProxy was able to significantly and consistently reduce 
the SLO violations in our experiments while reducing the cost of deployment by lowering the
number of required containers. \Cref{fig:all-res} shows the details on the way MLProxy is able to achieve such improvements. \Cref{fig:all-res-sub1} shows the 95th percentile of
response time throughout the experiment. According to the SLO in this experiment, our goal is to
keep the 95th percentile of the response time below 1000 ms. As shown in this figure, this value
for not using MLProxy is farther away from the SLO threshold, while MLProxy allows
the 95th percentile to stay closer to the SLO threshold to be able to batch requests together.

When the demand for our service (i.e., arrival rate) increases, MLProxy is able to allow a larger
maximum batch size (shown in \Cref{fig:all-res-sub4}) while still ensuring that the service abides
by the SLO. By leveraging the larger maximum batch size and due to the improved resource utilization 
made possible through batching, we can ensure an optimized deployment while maintaining a low SLO violation.

More details about the distribution of the response times can be seen in \Cref{fig:all-res-ccdf}.
As shown, MLProxy increases the latency for a portion of requests that would otherwise
be responded way faster than what is needed according to the SLO. However, this gives MLProxy
the flexibility needed to ensure a larger portion of requests is handled before the SLO
threshold.

\subsection{Resource Usage}

As can be seen in \Cref{tab:experiments}, MLProxy reduced the number of containers needed to
handle incoming requests between $32\%-92\%$ in our experiments. This improvement is a result of
utilizing the existing instances better and with lower overhead for each inference.
Due to the complex pricing schemas of Kubernetes offerings, we opted to use the container
count as a proxy to represent the deployment cost as it is often proportional to the cost of deployment
across different vendors.

It is worth mentioning that the MLProxy deployment used less than $200MB$ of memory and $10\%$ of a single
virtual CPU core in all experiments. As a result, the MLProxy deployment introduces negligible overhead
and can also be deployed by the cloud provider as an optional module on the API Gateway offerings.

\subsection{Discussions and Limitations}

In previous sections, we overviewed the experimental results of the proposed method. As shown, MLProxy
shows significant improvements in terms of both SLO violations and the cost of deployment. This can be
achieved by leveraging the flexibility allowed by the SLO and increasing the average response time. In general, the more flexible the SLO is, the
better the improvements would be when using MLProxy. However, these improvements mainly come from batching
the requests together when sending them to the upstream serverless platform. As a result, the benefits of
MLProxy are bound by the benefits achievable by batching the requests together. If, for a given workload,
batching doesn't reduce the overhead per request in a meaningful way, it wouldn't benefit from using
the proposed method due to the linear billing nature of serverless computing platforms.

Another effect of MLProxy on the system is reducing the frequency of change in the resources used by
a deployment. As a result, less resources will be dedicated to scaling the deployment, and this
can help us approach the performance of serverfull deployments while still getting the benefits of
serverless platforms.

As shown in \Cref{tab:workloads}, batching can reduce the overhead of the inference per request.
Other works in the field have also shown the benefits of batching requests in improving the device utilization
and cost~\cite{zhang2019mark,crankshaw2020inferline,crankshaw2017clipper,cox2020serverless} and
these benefits are expected to grow with the introduction of accelerated hardware into the systems.
As a result, the methodology proposed here would be even more beneficial in future serverless 
computing platforms.

While MLProxy addresses many concerns in machine learning inference serving on serverless computing
platforms, there are still a few limitations associated with it. As discussed earlier, the benefits
of using MLProxy mainly come from the benefits of batching requests together. Thus, for workloads where
the overhead reduction is negligible in larger batch sizes, MLProxy can't bring a lot of benefits.
This is because in serverless computing platforms, we have fine-grained pricing, and because of this,
a single instance running for $n$ request durations would cost the same as $n$ instances running for 1
request duration. Another limitation when using MLProxy is that to be able to achieve better performance
by batching requests together, the SLO should be flexible enough to allow a batch size of at least a few queries
without causing violations. In other words, the SLO threshold cannot be smaller than the 95th percentile
of the response time of the serverless platform when using a batch size of at least a few queries (e.g., five).

\section{Related Work} \label{sec:related-work}


Many studies have proposed methods to improve machine learning inference workloads on serverless computing platforms.
In \cite{crankshaw2020inferline}, The authors propose InferLine, which is a cost-aware optimizer that tries to find the optimal configuration to maintain a specified tail latency SLO by configuring the hardware type, batch size, and the number of replicas according to the model and arrival process. The approach taken by the authors is very promising but requires extra integration and control on the infrastructure and cannot function on serverless computing platforms.
Ali et al.~\cite{ali2020batch} evaluate adaptive batching for inference serving on serverless platforms. They used analytical performance models along with workload profiling and linear regression to find the optimal configuration for a given deployment. Although very promising, their work requires prior access to the deployed model and a profiling step that needs to be updated to be kept up to date. In addition, they do not investigate the performance of existing serverless systems and pure serverless model serving systems while only working with AWS Lambda.
Zhu et al.~\cite{zhu2019kelp} introduced Kelp, a software runtime that strives to isolate high-priority accelerated ML tasks from memory resource interference. They argue that in using accelerated machine learning, contention on host resources can significantly impact the efficiency of the accelerator. They show that their approach can improve the system efficiency by 17\%.
In \cite{zhang2019mark}, the authors propose MArk, a predictive resource autoscaling algorithm aiming to simplify machine learning inference and make it SLO-aware and cost-effective by combining IaaS, Spot Instances, and FaaS. In their design, they allow batching on accelerated deployments that use GPU/TPUs on IaaS to process large bulks of requests and use scale-per-request FaaS (AWS Lambda) as a tool to handle unforeseen surges in requests. In their approach, they were able to achieve an improved tail latency compared to state of the art in the industry while reducing the cost. The approach proposed in this work requires prior profiling steps that render it inefficient for serverless with pay-per-use pricing systems.
Crankshaw et al.~\cite{crankshaw2017clipper} proposed adaptive batching for serverfull deployments and were able to reduce the deployment cost and increase the throughput while meeting SLA.
Yadwadkar et al.~\cite{yadwadkar2019case} go over some of the challenges that arise for serving machine learning workloads like heterogeneous hardware and software, designing proper user interfaces, and building SLO-driven systems. In their work, the authors try to make a case for managed and model-less inference serving systems. Although non-trivial, our study could be a first step towards fully managed and cost-effective machine learning serving.
Chahal et al.~\cite{chahal2021performance} used a recommender system as an example
ML-based system to compare different deployment strategies on the cloud that result in
the desired performance at a minimal cost. In their experiments, they compared serverless and
serverfull deployments and found that serverless deployments deliver better performance
for bursty workloads and functions with short execution time and low resource requirements.
Wu et al.~\cite{wu2021serverless} investigated the possibility of using serverless computing
for ML serving and found that serverless computing platforms can deliver on
ML serving goals: high performance, low cost, and ease of management. They found small
memory size, limited running time, and lack of persistent state to be the most limiting factors
when deploying ML serving workloads on serverless computing platforms.
Benesova et al.~\cite{suppa-etal-2021-cost} explore the possibility of BERT-style text
analysis on serverless platforms. Their approximation methods allow the deployment of these
models on serverless platforms, eliminating the server management overhead and reducing the
deployment costs.
Gunasekaran et al.~\cite{gunasekaran2019spock} uses serverless computing alongside VM-based autoscaling with predictive and reactive controllers in order to improve SLO while reducing costs for machine learning inference workloads. In their work, they found that when the arrival rate is low, serverless computing platforms could be more cost-efficient that VMs because of their ability to scale to zero. Using our proposed platform, one can benefit the scale-to-zero capabilities of serverless computing while still having the ability to serve high-traffic workloads.
In previous studies, we have developed and evaluated steady-state and transient performance models
along with simulators for serverless computing platforms~\cite{mahmoudi2020tccserverless,mahmoudi2020tempperf,mahmoudi2021simfaas} with
homogeneous workloads. However, the unique characteristics and challenges in machine learning
inference workloads, along with the ever-lasting need for adaptive methods for optimization
components, led to the development of MLProxy.



Several of the recent studies investigated different methods to optimize machine learning serving workloads.
Romero et al.~\cite{romero2020infaas} proposed a model-less and managed inference
serving system. In this work, the authors generate model variants using layer fusion or
quantization to create models with varying performance/cost. They were able to find the best
combination of VMs to serve workloads using only high-level details about the Queries
Per Second (QPS), latency requirements, acceptable accuracy, and cost. This approach can
help improve serverfull deployments. However, it does not adapt to serverless environments
and cannot work with applications where approximate solutions are unacceptable.
Lwakatare et al.~\cite{lwakatare2020large} investigated the challenges faced for developing, deploying, and maintaining ML-based systems at large scales in the industry. They found several challenges according to adaptability, scalability, safety, and privacy. They found the most challenges currently faced to be in relation to adaptability and scalability.
In~\cite{gujarati2017swayam}, the authors introduce Swayam, an engine for distributed autoscaling to meet SLAs for machine learning inference. To improve resource utilization, the authors allow real-time and batch requests with different levels of SLA to enter the system and they use global state estimation from local data to drive the autoscaling algorithm.
In~\cite{cox2020serverless}, the authors go over the KFServing project that aims to allow scale-to-zero for machine learning serving workloads using Knative. Another benefit of using Kubernetes for serverless inference is found to be the ability to use GPUs while benefiting from autoscaling provided by serverless computing.

There are studies in the literature that attempted to use serverless computing to improve the performance and possibly cost of training machine learning models.
Feng et al.~\cite{feng2018exploring} investigated the challenges faced when using serverless computing platforms for training machine learning models. They found that serverless computing platforms could be leveraged for hyper-parameter tuning of smaller machine learning models to provide better parallelism. They found the most challenging issues of using serverless computing runtimes for training machine learning models to be their ephemerality, statelessness, and warm-up latency.
In~\cite{jiang2021towards}, Jiang et al. present a comparative study of distributed ML training over FaaS and IaaS. They found that serverless training is only cost-effective with models that have a reduced communication overhead and quick convergence.



Other studies have focused on investigating the challenges and opportunities of different paradigms in cloud computing for inference in machine learning systems.
Lwakatare et al.~\cite{lwakatare2019taxonomy} investigated the most challenging aspects of integrating
machine learning systems into software products and found dataset assembly and model creation,
training, evaluation, and deployment to be the most important ones.
Elordi et al.~\cite{elordi2020benchmarking} evaluated deployment of Deep Neural Network (DNN) models on a serverless
computing platform like AWS Lambda. They used common workloads used in MLPerf~\cite{reddi2020mlperf} and
found that increasing the memory of the serverless deployment has a huge impact on the response
time but a less dramatic effect on the total throughput of the system. In their experiments,
they found the serverless computing platform to be capable of making 51-83 inferences per second,
making serverless a suitable deployment point for DNN workloads.

In this study, our main focus has been on analyzing and improving the performance
and cost of machine learning inference workloads on serverless computing platforms. Previous studies
have focused on other paradigms in cloud computing, but serverless computing has been shown
to suffer from bad cost-performance tradeoffs. Our goal here has been to help mitigate this
effect for machine learning inference workloads, making serverless computing a viable
option for these deployments.


\section{Conclusion} \label{sec:conc}

In this work, we presented and evaluated MLProxy, which is an adaptive reverse proxy to support
efficient and SLO-aware batching for machine learning inference serving types of workloads.
We analyzed and evaluated both deployment cost and performance implications of MLProxy in
current serverless computing platforms and we showed the effectiveness of the proposed method
through extensive experimentation. We also showed that MLProxy can work across different machine learning
libraries. The proposed system can be used by serverless computing providers as a part
of their API Gateway offering, significantly improving their system efficiency and competitive
advantage. It can also be deployed by the developers in a Kubernetes cluster supporting their serverless
deployment with minimal added cost due to the low resource needs of the system.

In summary, the proposed reverse proxy can help developers use the deployed resources in their
serverless deployments more efficiently while maintaining the SLO requirements of
their machine learning serving system.

\section*{Acknowledgements}
This research was enabled in part by support from Sharcnet (www. sharcnet.ca) and Compute Canada (www.computecanada.ca).
We would like to thank Cybera, Alberta's not-for-profit technology accelerator, who supports this 
research through its Rapid Access Cloud services.


\bibliographystyle{ACM-Reference-Format}
\bibliography{bibliography}

\end{document}